# Linear complexions directly modify dislocation motion in face-centered cubic alloys

Divya Singh [a], Vladyslav Turlo [b,d], Daniel S. Gianola [c], Timothy J. Rupert [a,d,*]

[a] Department of Materials Science and Engineering, University of California, Irvine, CA 92697, USA
[b] Laboratory for Advanced Materials Processing (LAMP), Swiss Federal Laboratories for Materials Science and Technology (Empa), Thun, CH-3602, Switzerland
[c] Materials Department, University of California, Santa Barbara, CA 93106, USA
[d] Department of Mechanical and Aerospace Engineering, University of California, Irvine, CA 92697, USA
*Email: trupert@uci.edu

**Abstract**

Linear complexions are defect phases that form in the presence of dislocations and thus are promising for the direct control of plasticity.  In this study, atomistic simulations are used to model the effect of linear complexions on dislocation-based mechanisms for plasticity, demonstrating unique behaviors that differ from classical dislocation glide mechanisms.  Linear complexions impart higher resistance to the initiation and continuation of dislocation motion when compared to solid solution strengthening in all of the face-centered cubic alloys investigated here, with the exact strengthening level determined by the linear complexion type.  Stacking fault linear complexions impart the most pronounced strengthening effect, as the dislocation core is delocalized, and initiation of plastic flow requires a dislocation nucleation event.  The nanoparticle and platelet array linear complexions impart strengthening by acting as pinning sites for the dislocations, where the dislocations unpin one at a time through bowing mechanisms.  For the nanoparticle arrays, this event occurs even though the obstacles do not cross the slip plane and instead only interact through modification of the dislocation's stress field.  The bowing modes observed in the current work appear similar to traditional Orowan bowing around classical precipitates but differ in a number of important ways depending on the complexion type.  As a whole, this study demonstrates that linear complexions are a unique tool for microstructure engineering that can allow for the creation of alloys with new plastic deformation mechanisms and extreme strength.

# 1. Introduction

Complexions are thermodynamically-stable chemical and/or structural states confined to defects such as grain boundaries and dislocations [1-9]. As linear complexions (one-dimensional defect phases confined to the dislocation lines) were discovered only recently, most of the existing literature is concentrated on grain boundary complexions [3,4, 10-20]. A complexion transition is typically limited to localized regions near the defect that provides a template for solute segregation, and grain boundaries can go through structural transitions similar to bulk phase transitions due to variation in parameters such as temperature or local composition [14, 19, 21-24]. Such abrupt changes in the structure, composition, stress, and local properties of the grain boundary can lead to either beneficial or detrimental changes in material response including atom transport properties [14], mechanical properties [20, 25-28] and thermal stability of polycrystals [20-21, 28-30]. For instance, annealing experiments show that increased high temperature stability of nanocrystalline face-centered cubic Ni- and Cu-rich alloys can result from the presence of amorphous intergranular complexions [20, 21]. In addition, nanocrystalline Cu-Zr can be strengthened by disordered grain boundary complexions [21], which was shown to result from stronger dislocation pinning that increased the flow stresses needed for dislocation propagation [25]. There is also evidence that combining nanostructuring with complexion engineering can increase ductility without the typical reduction in strength [27, 28]. In contrast, polycrystalline Ni samples doped with Si showed brittle intergranular fracture due to the formation of bipolar interfacial structures within grain boundary complexions [26]. These bipolar structures were shown to make it easier for grain boundary fracture/decohesion to occur, explaining the overall embrittlement effect.

Strengthening of metal alloys generally comes from the retardation of dislocation glide, with traditional precipitates or grain boundary complexions being static features of the microstructure than can act as obstacles to dislocations. Their effect is therefore indirect on plasticity and strength. However, dislocations themselves have local variations in structure and stress state and thus they can potentially host complexions as well, which would open a new pathway for strength modification through the manipulation of the direct carriers of plasticity. Kuzmina et al. [1] first reported the existence of linear complexions by carefully investigating the local structure near dislocations in an Fe-9 at.% Mn alloy. These authors showed that face-centered cubic regions with an austenitic composition could form and be in equilibrium with the body-centered cubic matrix, but that these complexions were limited to the areas near the dislocations. Modification of a dislocation's environment should impact plasticity, and in fact static strain aging was reported in a follow-up study by Kwiatowski da Silva et al. [31]. Additional evidence of linear complexions in body-centered cubic metals was provided by Odette et al., who observed the formation of stable precipitates along dislocation networks in irradiated low alloy reactor pressure vessel steels [32]. Turlo and Rupert [8] isolated the atomic-scale mechanisms responsible for strengthening due to linear complexions in body-centered cubic alloys by performing atomistic modeling of dislocation break-away in an Fe-Ni alloy. These authors reported that the complexion's pinning effect was related to a reduction of the dislocation's intrinsic stress field and that a diffusion-less, lattice distortive transformation of the complexion can occur once the dislocation is able to move away.

A number of linear complexions in face-centered cubic alloys were recently predicted by Turlo and Rupert [9] using atomistic simulations. Three types of linear complexions were found and classified according to their effect on the original dislocation core: (1) Nanoparticle array linear complexions, (2) platelet array linear complexions, and (3) stacking fault linear complexions. Nanoparticle array linear complexions were predicted for Ni-Fe, Ni-Al and Al-Zr alloys, where arrays of L1$_2$ phase intermetallic particles are hosted due to stress-driven segregation, with the dislocation cores remaining unaltered. Platelet array linear complexions were predicted for Al-Cu, where Cu segregates and then forms disc-shaped platelets that grow out of the slip plane and along the dislocation length. These complexion platelets have the same structure as Guinier-Preston (GP) zones observed during the aging of Al-Cu bulk alloys [33-42]. For stacking fault linear complexions, Zr segregates to both the partial dislocations and the stacking fault itself, forming Cu$_5$Zr intermetallic at elevated concentrations. This particular linear complexion type delocalizes the core of the original dislocation. With a delocalized core, stacking fault linear complexions can have a thickness normal to the original slip plane that is more than one atomic spacing. Zhou et al. [43] recently found experimental evidence of ~~nanoparticle array linear~~ more complexed and varied solute segregation in a face-centered cubic Pt-Au alloy, with distinct chemical segregation profiles observed across different defects comprised of dislocation networks such as bulk glissile dislocations, low-angle grain boundaries, stacking fault tetrahedra, and Frank loops. This segregation was of a greater magnitude than prior observations, such as the well-known Cottrell atmosphere in body-centered cubic metals, and also occurs by segregation of solute atoms, rather than smaller impurity atoms. Ref. [43] found that the local segregation was closely dependent on the defect type as well as the exact topology of the defect, suggesting a connection of local composition with the stress fields associated with the defects. This observation is indicative of close relation between the stress fields associated with the defect and the level of Au segregation facilitated by it. The observation of defect-specific solute concentrations by these authors opens the door for a materials design approach to defect engineering, where local structure and chemistry can be planned. Linear complexions, as states that are localized to the dislocation region, are promising for the development of novel high strength face-centered cubic alloys.

Intuitively, the dislocation mechanics, strengthening mechanisms, and evolution of linear complexion under mechanical stress in face-centered cubic alloys should be intimately related to the complexion type and structural descriptors. However, the impact of linear complexions on the yield strength of face-centered cubic metals and the extent of strengthening increments that can be achieved have not been investigated to date.

In this work, we provide a comprehensive computational study of dislocation motion in the presence of the three types of linear complexions uncovered to date in face-centered cubic metals, isolating their effect on mechanical behavior. All of the samples with linear complexions exhibit much higher strengths when compared to random solid solutions of the same composition, with the strengthening effect dependent on complexion type. The stresses required for successive depinning of dislocations from nanoscale precipitate arrays is sensitive to the particle size and their relative distribution along the dislocation line. Platelet array linear complexions push the partial dislocations out of their slip planes, with these segments acting as pinning points for the dislocations and overlapping complexion platelets displaying the highest pinning capacity. Stacking fault linear complexions delocalize the dislocation core which necessitates the nucleation of a new dislocation, sometimes altering the dislocation's slip plane, and results in the most pronounced strengthening effect. Overall, this work demonstrates new avenues for the manipulation of mechanical behavior through defect engineering.

## 2. Methods

All simulations were carried out using the Large-scale Atomic/Molecular Massively Parallel Simulator (LAMMPS) [44] and many-body interatomic potentials following the embedded atom method (EAM) formalism for Cu-Zr [45], Ni-Al [46], Al-Cu and Al-Cu-Zr [47].

The interatomic potentials used in this study were selected to reproduce important features of the experimental phase diagrams such as the major stable phases and solubility limits, which is crucial to capturing the structure and chemistry of nanoscale linear complexions. The potentials for Cu-Zr, Al-Zr and Al-Cu have also been previously employed to study grain boundary complexion transitions [23, 25, 48-49]. Atomic scale analysis and visualization was performed with OVITO [50] using the Polyhedral Template Matching (PTM) [51] and Dislocation Extraction Algorithm (DXA) [52] methods. PTM was used to analyze the local atomic ordering while DXA was used to analyze dislocation structure. The coloring scheme employed in the work is as follows: Face-centered cubic atoms are green, body-centered cubic atoms are blue, hexagonal close packed atoms are red, icosahedral atoms are yellow, and the $L1_2$ intermetallic phase is denoted by magenta color. Any atoms that cannot be categorized according to these crystallographic structures (i.e., are identified as "other" atoms) are represented by white coloring.

The initial sample geometry is shown in Figure 1(a), where a pair of edge dislocations have been inserted into the middle of the sample. The initial edge dislocation pair relaxes into two Shockley partial dislocations separated by a stacking fault with the application of conjugate gradient energy minimization. Each simulation cell was constructed as $60\sqrt{2}\ a$ in the X-direction, $48\sqrt{3}\ a$ in the Y-direction and $50\sqrt{6}\ a$ in the Z-direction, where $a$ is the lattice constant of the host atom of the alloy system. Typical sample dimensions are presented in Figure 1(a) along the edges of the simulation cell, although the exact sizes of the simulation boxes vary slightly depending on the lattice constant of each material at a given temperature and composition. The simulation cells contained ~3,500,000 atoms, representing very large atomistic models (see, e.g., [53] for comparison). Periodic boundary conditions were employed in all directions. Two types of samples for each alloy system are considered for a given composition: (1) A random solid solution and (2) an equilibrated state where dopant atoms have segregated to form linear complexions. Three thermodynamically equivalent simulation cells (different atomic vibrations but all other thermodynamic parameters held constant) were studied for each sample type to enable improved statistics.

The equilibrium states of linear complexion samples were obtained using the hybrid Monte Carlo (MC)/molecular dynamics (MD) method, with MC performed following the variance-constrained semi-grand canonical ensemble proposed by Sadigh et al. [54]. Hybrid MC/MD simulations allow for both chemical segregation and local structure relaxation, which enables local transformations such as the formation and growth of linear complexions to occur. In this study, one MC step was performed for every 100 MD time steps. The system was considered to have equilibrated when the energy gradient in last 20 ps of MD simulation was less than 0.1 eV/ps. Additional details of the MC/MD of simulation procedure can be found in Ref. [9]. Both the solid solution and linear complexion samples were next equilibrated at 300 K and zero pressure under an NPT ensemble. Nanoparticle array linear complexions are shown in Figure 1(b) and outlined in red, platelet array linear complexions are shown in Figure 1(c) and outlined in blue, and a stacking fault linear complexion is presented in Figure 1(d) and outlined in green. These colors are used in subsequent figures to denote the complexion type. Platelet array complexion restructure the dislocation core by creating facets along the line direction, while stacking fault complexions delocalize the dislocation core. The Al-Cu-Zr system has both platelets and nanoparticles in the same locations, and therefore have a coexistence of two complexion types. All simulation cells contain multiple linear complexion particles along the dislocation line length, ensuring that a population of obstacles is probed during deformation. Mechanical testing consisted of the application of shear strain with a strain rate of $10^8$ s$^{-1}$ at 300 K. Shear stress-strain curves were plotted for each sample and each simulation run. We note that the linear complexions are only truly stable in the presence of the dislocations, as their stress-fields drive segregation and any associated structural transitions. This means that a linear complexion may evolve or remain in a metastable configuration after the dislocations have moved away, with the exact diffusion mechanism and time scales involved likely dependent on the complexion type, alloy choice, and testing temperature. To avoid any such uncertainty, we focus our analysis and discussion on the initial yield event, where the dislocations are attempting to move away from the equilibrium configuration.

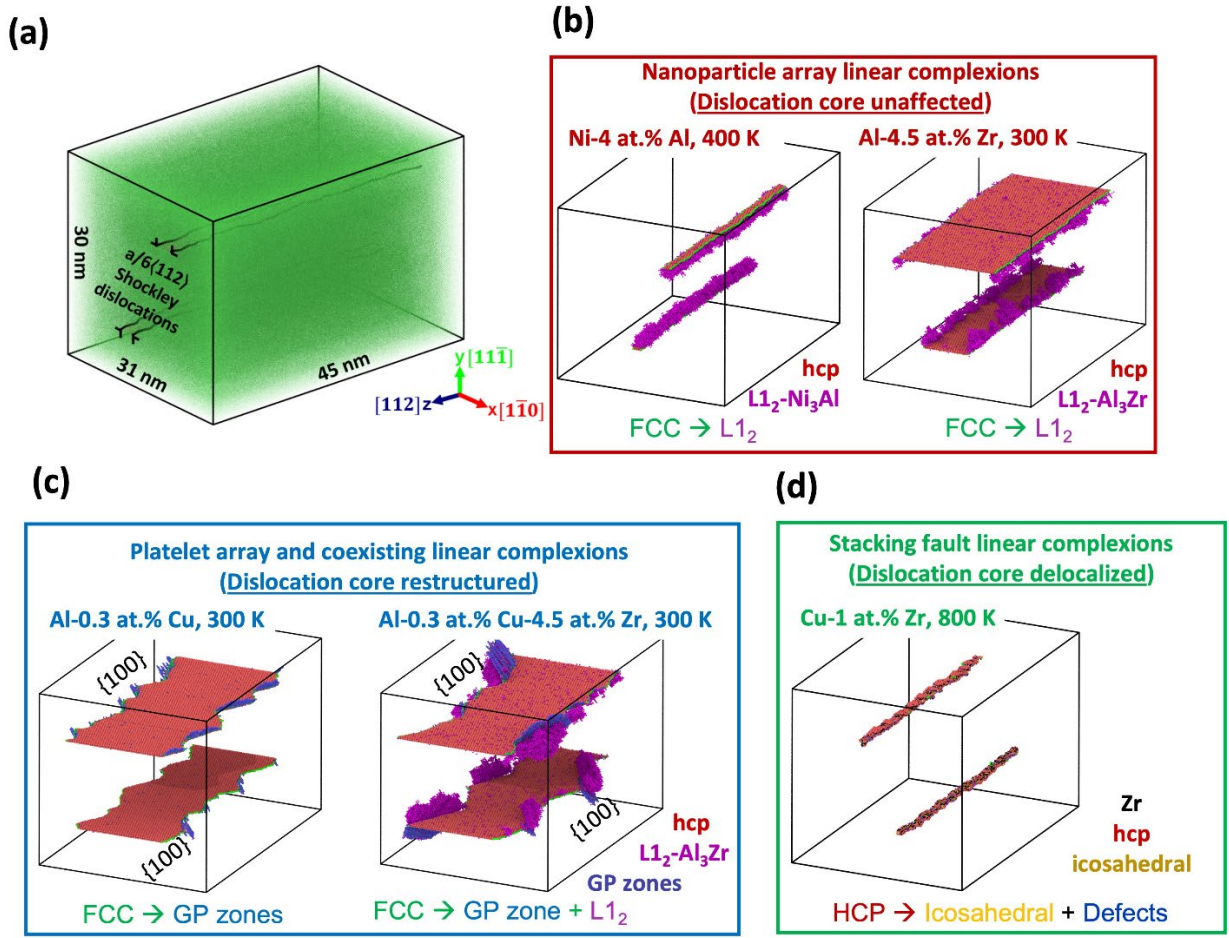

**Figure 1.** The simulation cell with two dislocations split into two Shockley partials and a stacking fault between them, (a) before and (b-d) after equilibration at the listed compositions and temperatures.

# 3. Results

## 3.1. Breakaway stress and flow stress measurements

The shear stress-strain curves obtained from all five alloys, in both solid solution and linear complexion configurations, are presented in Figure 2. The noticeable variations in the stress-strain curves demonstrate that solutes and linear complexions interact with dislocations in different ways. In order to compare the two situations in a more quantitative manner, it is useful to define critical points on the stress-strain curves. The first sudden drop in the stress can be termed the ***breakaway stress***, as it marks the start of dislocation glide after first depinning from the local solute or complexion obstacles. The breakaway stress is analogous to the yield stress of the chosen simulation cell. Subsequent stress drops that appear on the stress-strain curves after the breakaway point are analogous to flow stresses, as these are the values needed to continue plastic deformation of the sample. These events are related to the dislocation leaving and returning through the periodic boundaries and are termed ***peak stresses***. Examples of both critical stress quantities are shown in Figure 3(a) for an example curve in Ni-Al.

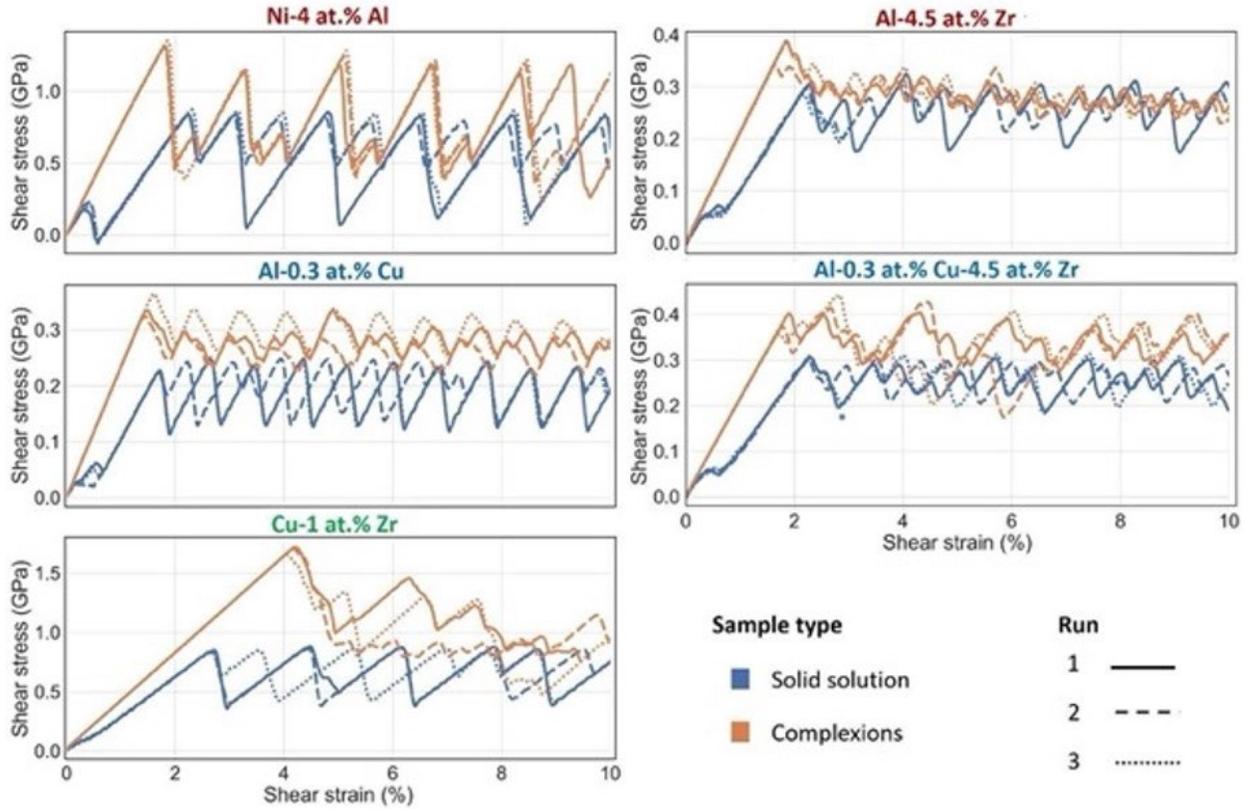

**Figure 2. Stress-strain curves for different samples with linear complexion (orange) and for a solid solution (blue). Three thermodynamically equivalent but microscopically different samples are tested for statistical purposes and shown in different line styles.**

To allow for comparison between different base metals, normalized versions of the two critical stresses are obtained by dividing the absolute stress values by the shear modulus, $G$, of the base metal, with these metrics plotted in Figures 3(b) and (c). The normalized values of peak stresses are calculated by averaging over the different peak stresses measured for the entire simulation run. This approach is similar to comparing the Peierls barrier of different metals in terms of normalized Peierls stress [55], rather than comparing their absolute values. For example, Figure 3(b) shows that the values of normalized breakaway stresses are of similar magnitude for Ni-4 at.% Al, Al-4.5 at.% Zr, Al-0.3 at.% Cu and Al-0.3 at.% Cu-4.5 at.% Zr samples with linear complexions, while the breakaway stress for the Cu-1 at.% Zr sample is much higher. It is noteworthy that even though defects are present, the Cu-1 at.% Zr sample shows a breakaway stress ~0.04G that is a significant fraction of the theoretical shear strength of Cu, previously estimated to be ~0.07G for a defect-free crystal [56]. In the subsequent sections, the unique dislocation mechanisms associated with critical points on the stress-strain curves are isolated for different types of linear complexions. We focus our attention primarily on the initial breakaway event, as this is the critical first plastic event when the system evolves away from the equilibrated initial state. Again, this is to avoid any uncertainty about the structure of the linear complexion particles after the host dislocations have moved away

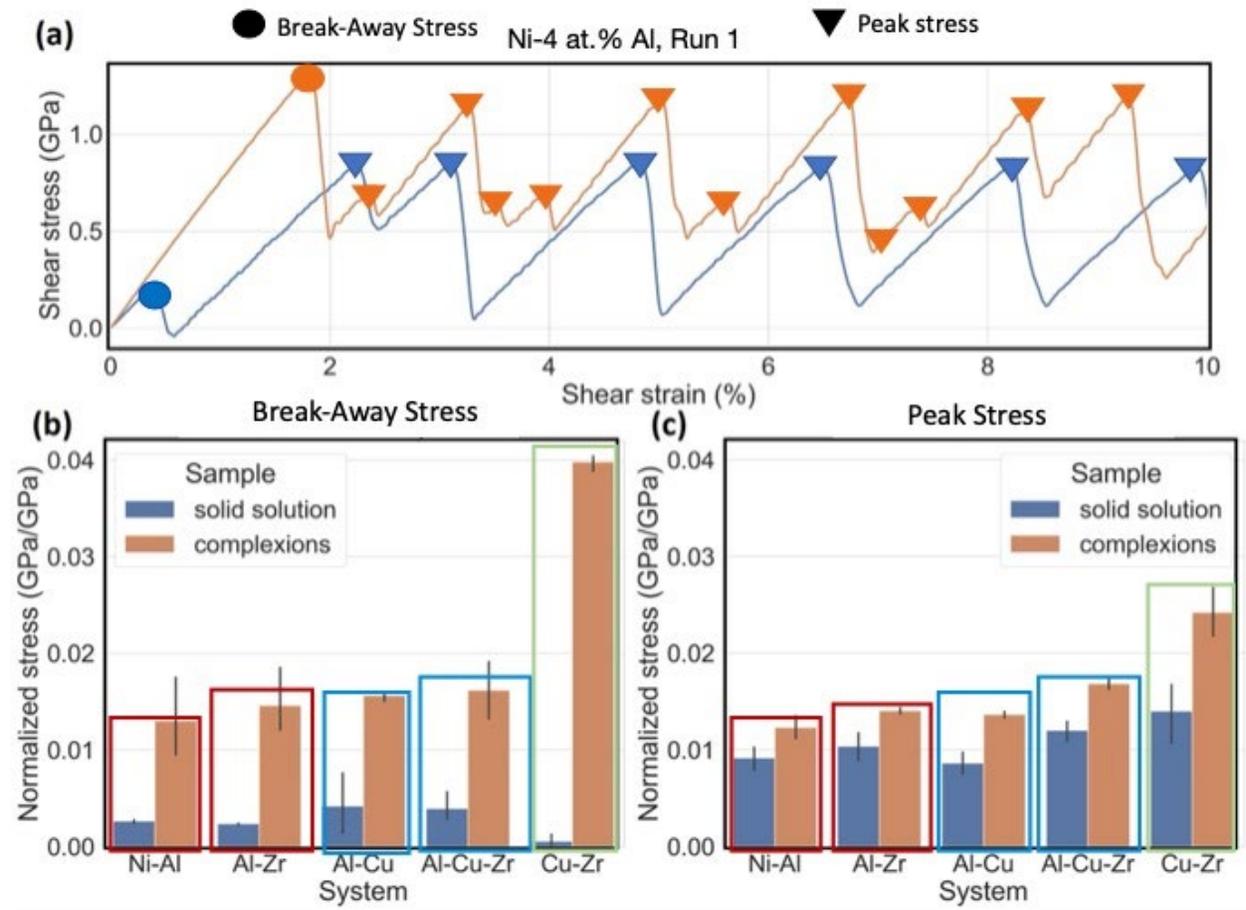

**Figure 3. (a) Representative shear stress-strain taken from the Ni-Al sample, to show the definition of breakaway and peak stresses. The quantified metrics for different alloys and sample types are presented for (b) breakaway stress and (c) peak stresses, which are analogous to yield strength and flow stress, respectively. Error bars are standard deviations obtained from the different runs.**

## 3.2. Dislocation mechanics - Nanoparticle array linear complexion

Figure 4(a) isolates the initial dislocation breakaway event from a complexion in Ni-4 at.% Al, to serve as a representative example of a nanoparticle array linear complexion. A zoomed view of the stress-strain response is shown near the breakaway event, with critical events A-F marked on the curve and also shown in detail on the right side of Figure 4(a). The evolution of dislocation line length and stacking fault size (presented as the total number of atoms in the stacking fault, which is proportional to the stacking fault width since the simulation cell has a constant length along the direction of the dislocation line direction) are also plotted in this figure. First, it is notable that the particles are not on the glide plane of the dislocations, but on the tension side of the dislocation stress field. This means the particles are located either above or below the glide plane, depending on which dislocation pair one looks at (top or bottom, which are originally positive and negative edge dislocations, respectively). The linear complexions interact with the dislocations through their stress fields, rather than blocking the glide path directly. The dislocations remain pinned by the complexion up to an applied shear strain of ~1.6%, corresponding to panel B, where the maximum stress is reached. One can also observe an increase in the stacking fault size, as the leading partial in the top pair begins to pull away while the trailing partial dislocation remains fixed. This is followed quickly by the first full unpinning event of the top dislocation pair, shown in panel C, where a bowing motion is observed as the defect unpins from the complexion nanoparticles sequentially. The regions near the front and back of the simulation cell have not broken away yet. The increase in the curvature of the dislocation due to bowing appears as an increase in the dislocation line length. The second pair of partial dislocations in the bottom of the simulation cell also begins to move at this point, unpinning through a similar motion yet moving in the opposite direction because it is a dislocation of the opposite sign. Event D shows that both dislocation pairs have moved away from the complexion and are traveling through the clean areas of the simulation cell. A corresponding decrease in the stacking fault size (i.e., contraction of the stacking fault width) is observed here, as the partial dislocation pairs are outside the influence of the complexion particle at this stage. The dislocations continue to glide until they reach the periodic boundaries, which allow the dislocations to re-enter the simulation cell from the other side and eventually come back to their original positions above/below the linear complexion particles (panels E and F), at which point the dislocation line length and stacking fault size return to the initial values. The dislocations are again pinned due to a reduction of their stress fields in the presence of the particles. During this sequence, the partial dislocation pairs generally move together, likely due to the fact that there is only one array of complexion particles for each pair and therefore a common pinning obstacle.

A similar sequence of bowing and breakaway events is observed for all alloys which exhibited the nanoparticle array linear complexion type, although important differences can occur, with another example shown for Al-4 at.% Zr in Figure 4(b). A two-dimensional projection viewing the top half of simulation cell along the Y-axis is shown, with only one set of partial dislocations and their accompanying complexion particles appearing. The partial dislocation on the left (the trailing partial) experiences bowing and unpinning first, in the top half of the images. This trailing partial dislocation crosses the gap between particles and eventually pushes the second, leading partial dislocation on the right until it breaks away at ~1.88% shear strain. The first particle then becomes pinned at the second array of complexion nanoparticles. Interestingly, the segment of the partial dislocation pair in the top of Figure 4(b) undergoes this breakaway and re-pinning sequence before the bottom half of the same partial pair has moved away from its host particles. These observations highlight the fact that the partial dislocations can move independently, with this effect likely being most pronounced when each partial dislocation has nucleated its own nanoparticle array.

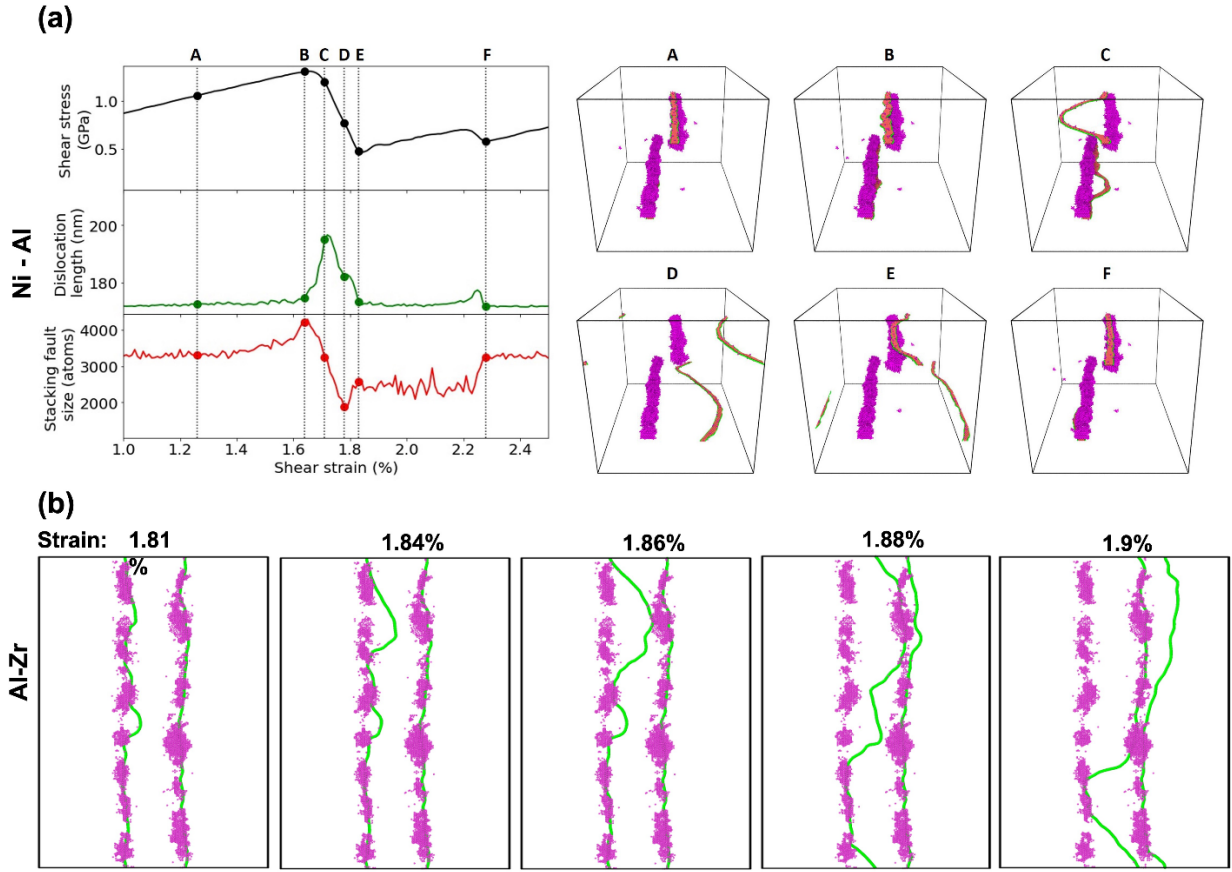

**Figure 4. (a) The evolution of shear stress, dislocation length, and stacking fault size as a function of shear strain for the breakaway stress events in a Ni-4 at.% Al alloy with nanoparticle array linear complexions, with corresponding snapshots shown to the right. (b) Similar unpinning and bowing events for one pair of partial dislocations on the same slip plane in Al-4 at.% Zr, with the applied shear strain labeled at the top of the images.**

An important common observation for nanoparticle array linear complexions is that one dislocation segment (whether an individual partial dislocation for Al-Zr or a pair of partials for Ni-Al) is mobile, while the other dislocation segments remain static for a time. This response demonstrates a non-identical "pinning power" of similar, but not identical, arrays of complexion along the dislocations. For example, Figure 4(b) shows that two small bowing events are observed at gaps in the particles in the panel corresponding to 1.84% shear strain, suggesting that such particle-free regions can be weak links. This behavior can be understood by recalling that the complexion particles form to reduce the dislocation's stress field and lower the overall system energy, meaning that the region near the dislocation with elevated hydrostatic stress is a roughly cylindrical template above or below the defect (depending on whether the original dislocation is of positive or negative edge character) that can be occupied by obstacles to motion. Complexion particles pin the dislocation in a configuration where it is energetically preferred while the lack of a particle corresponds to a dislocation segment that is not effectively pinned. To investigate this effect more clearly, additional models of nanoparticle array linear complexions were created in the Ni-Al system with varying Al concentrations, which allows the density of linear complexion particles to be systematically increased (Figure 5(a)). The particles grow to more fully decorate or fill the roughly cylindrical regions below/above the dislocation line as Al concentration is increased, leading to fewer and smaller gaps between the particles. Shear deformation simulations in Figure 5(b) show that the breakaway stress increases with Al concentration as well, suggesting that the closure of the gaps between particles increases the strength of the pinning effect. These effects will be compared to classical strengthening models later.

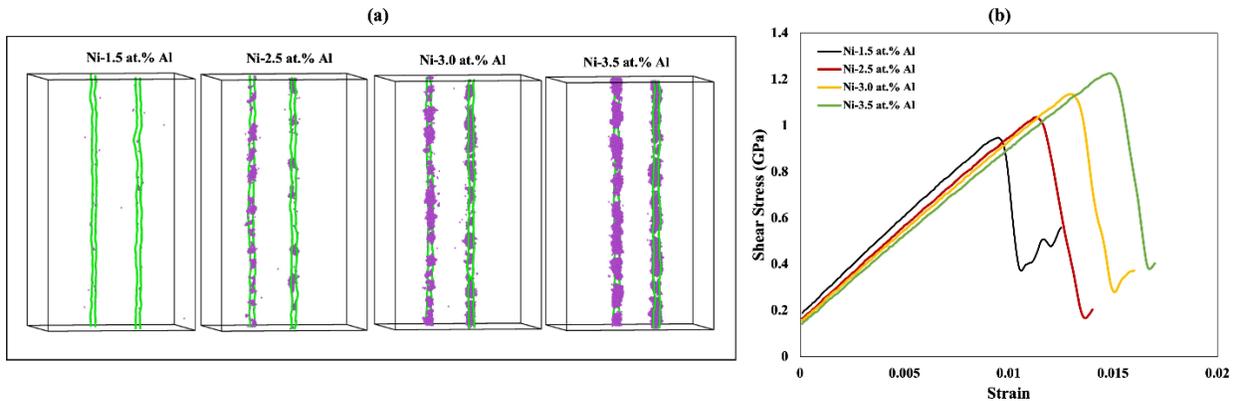

**Figure 5. (a) Nanoparticle array linear complexions for Ni-Al with increasing Al concentration and therefore more complexion particles. (b) Shear stress-strain curves show that the breakaway stress is significantly increased as the complexion particles become more densely populated along the dislocation line.**

### 3.3. Dislocation mechanics – Platelet array linear complexion

The effect of platelet array linear complexions on dislocation mechanics and strengthening is next explored in the context of an Al-Cu alloy. The stress-strain response as well as the evolution of dislocation line length and stacking fault size is presented in Figure 6 for the initial breakaway event (bottom partial dislocation pair), with important events again called out in panels A-F. The trailing partial dislocation on the left first bows in a region where there are no GP zones (denoted by a blue arrow in panel B) and then unpins from the platelets one by one, while the leading partial dislocation on the right remains stationary. This behavior reveals that the strengthening provided by the platelet complexion arrays is also dependent on their local arrangement, both in terms of platelet size and the gaps between them. As a result of slip events B-C in Figure 5, a decrease in the stacking fault width is measured. Finally, the gliding trailing partial interacts with the stationary leading partial, eventually pushing the leading partial to move and continue plastic flow. Panel D shows the glide of the leading partial dislocation across the periodic boundary conditions and back into the other side of the simulation cell. It is interesting to note that the leading partial does not interact strongly with the existing GP zones platelets on the left side of the cell and instead continues to glide relatively unhindered through the crystal in panels D-E. Finally, the partials return to their original locations along the platelet complexion arrays in panel F. Both the dislocation length and stacking fault size first decrease and then increase during this breakaway event sequence, returning to the initial values at the end. Subsequent plastic flow events have identical fluctuations in the dislocation line length and stacking fault size, demonstrating that the same mechanism operates again and again during the simulated deformation.

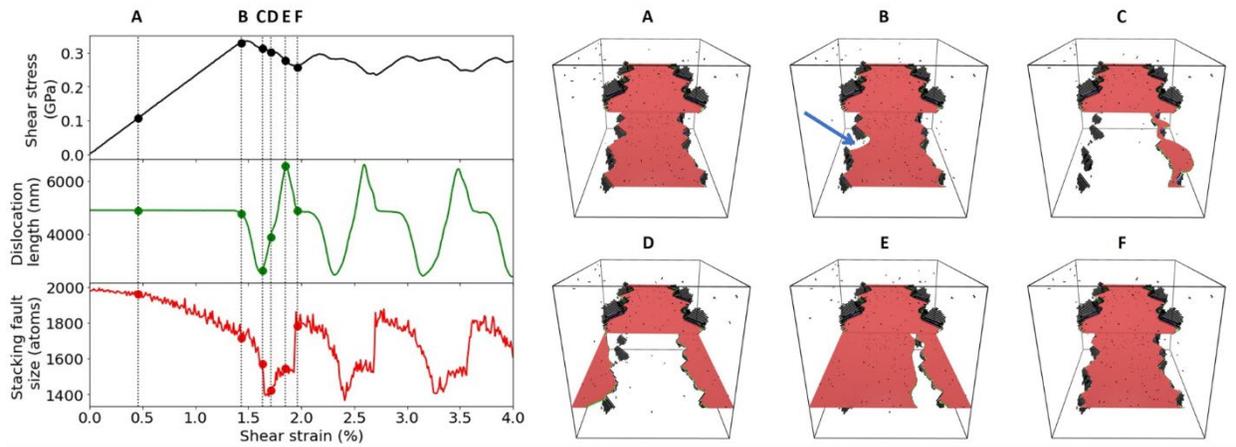

**Figure 6. The evolution of stress, total dislocation length, and stacking fault size are shown for the first unpinning and pinning events, with atomic snapshots corresponding to important events, for the platelet array linear complexions in the Al-0.3 at.% Cu system.**

In general terms, the deformation mechanisms for the platelet array linear complexions resemble those of the nanoparticle array complexions, as the dislocations begin to bow in regions without complexion platelets/particles until the entire line can move away from its pinning points. However, critical differences can be observed, particularly with regards to the importance of the GP zone orientations. While the nanoparticle arrays did not alter the dislocation core, the platelet arrays cause a faceting of the dislocation lines related to the crystallography of the GP zones. Figure 7 shows a detailed view of different pinning events, where a strong pinning event (outlined in a dashed blue box) can be seen at 1.55% shear strain. The zoomed view in Figure 7(b1) shows that the trailing partial has formed a segment with edge character along the GP zone, but also that this segment has shifted up along the GP zone to be slightly out of the original glide plane. This faceted, out-of-plane feature creates a strong obstacle to motion, with pronounced bowing of the red screw segments needed to eventually break free. Figures 7(b2) and (b3) show that a similar faceting occurs for the leading partial and also that small segments of the two partials can recombine. The subsequent flow events are also determined by the local platelet and dislocation configurations. Figures 7(d1) and (d2) show the strongest pinning site for the follow-up flow event, where again recombination of the partials occurs. In addition, there are overlapping GP zones in this region, which result in a very complicated faceted structure that must be overcome.

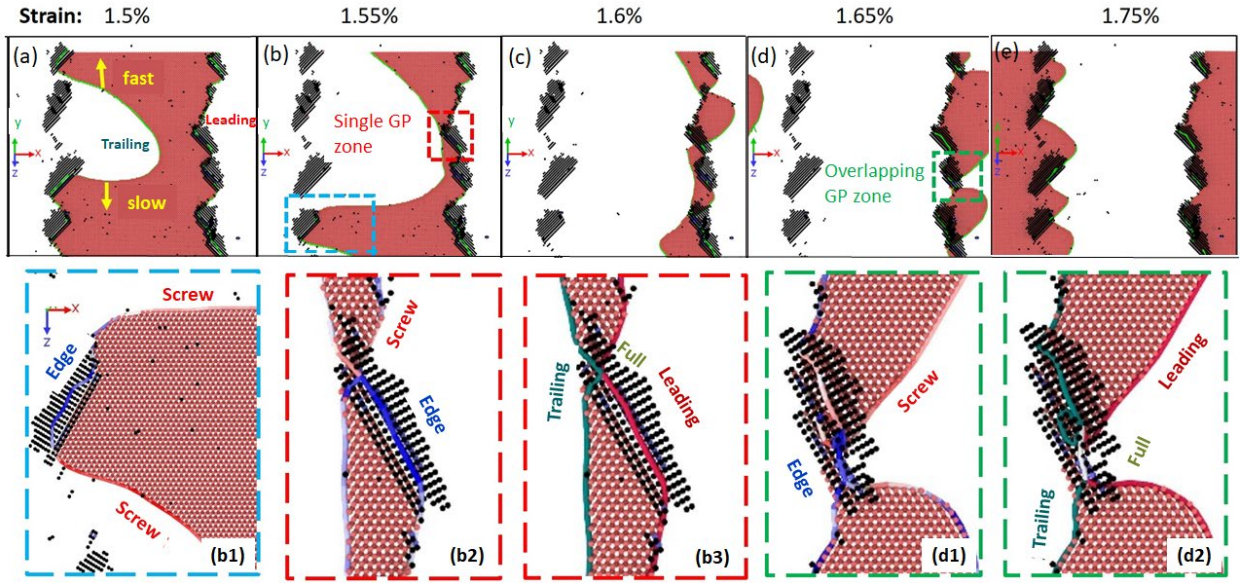

**Figure 7.** (a)-(e) Partial dislocation interactions with the GP zones comprising the platelet array linear complexion in the Al-Cu alloy. (b1) The trailing partial remains pinned due to a segment with edge character that has moved slightly out-of-plane along the GP zone. (b2)-(b3) The trailing and leading partials can recombine at some small sites, while edge and screw character segments also form. (d1)-(d2) A strong pinning event occurs at a region with overlapping GP zones, where partial dislocation recombination also occurs.

### 3.4. Dislocation mechanics - Stacking fault type linear complexion

Stacking fault type linear complexions delocalize the dislocation core into a broader region. The changes in the shear stress, dislocation line length, and stacking fault size are presented for the first plastic event in Figure 8 as a function of shear strain. A high breakaway stress of 1.5 GPa is recorded at an applied shear strain of 4.2%, at which point the nucleation of a new partial dislocation occurs from the delocalized dislocation core in the top half of the simulation cell (panel B). Once nucleation occurs, the new segment grows and an increase in dislocation line length can be seen. The leading partial dislocation can quickly move through the simulation cell, leaving a stacking fault behind it and resulting in a rapid increase in the stacking fault size. The dislocation re-enters the simulation cell due to the periodic boundary conditions and begins to be re-absorbed into the complexion in panel D1, where the beginning stages of nucleation for the trailing partial dislocation are also observed. Eventually the trailing partial also moves out and then back into the simulation cell to be re-absorbed. The bottom complexion undergoes a similar process, with the leading partial being nucleated at a larger applied strain in panel C. By panel F, the leading partial has been re-absorbed while the trailing partial has not been nucleated yet, leaving a stacking fault that crosses the entire simulation cell for the time being. The nucleation of partials at different stages of the simulated deformation again emphasizes that local variations in complexion structure determine the individual breakaway events and their associated barriers.

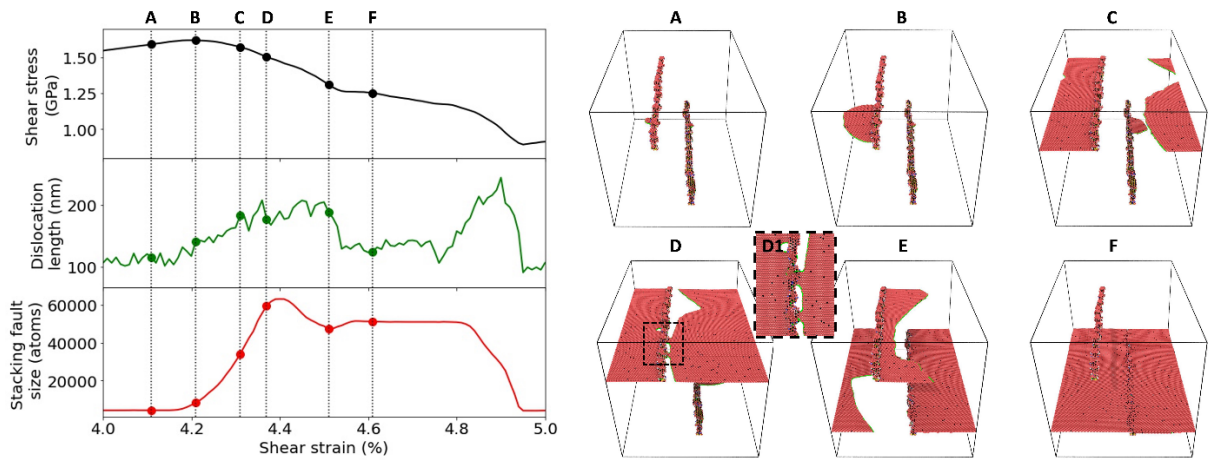

**Figure 8. The evolution of stress, total dislocation length, and stacking fault size are shown for the first unpinning and pinning events, with atomic snapshots corresponding to important events, for the platelet array linear complexions in the Cu-1 at.% Zr system.**

The nucleation event for stacking fault linear complexions is shown in more detail in Figure 9. Figures 9(a) gives a top view, looking down onto the glide plane, of the delocalized dislocation core that first nucleated a partial in Figure 8. The dislocation extraction algorithm (DXA) in OVITO no longer identifies a continuous dislocation along the linear complexion, but instead finds an array of disconnected full edge and partial mixed character dislocations. Such dissociated and disconnected dislocation structures have also been measured in simulations of grain boundary structure [25, 57-70]. At 1% applied shear strain, the early stages of partial dislocation nucleation can be observed from one of the disconnected regions near the top of the image. As the strain increases and the leading partial moves away from the original location, one can observe that the region remaining shifts from being largely blue (full dislocation with edge character) to green (partial dislocation with mixed character). Eventually the entire leading partial has been nucleated and travels through the simulation cell, leaving another disconnected collection of defects. Figures 9(b) and (c) provide front views, down the original dislocation's line direction, for the solid solution and linear complexion version of the Cu-Zr alloy, respectively, where the original slip planes are marked by dashed black lines. For the solid solution, the dislocations glide easily on the original slip planes. In contrast, Figure 9(c) shows that the delocalization of the dislocation core also occurs normal to the slip plane, which can lead to nucleation on a different but parallel slip plane as deformation continues. We note that 4.3% shear strain corresponds to panel C in Figure 8.

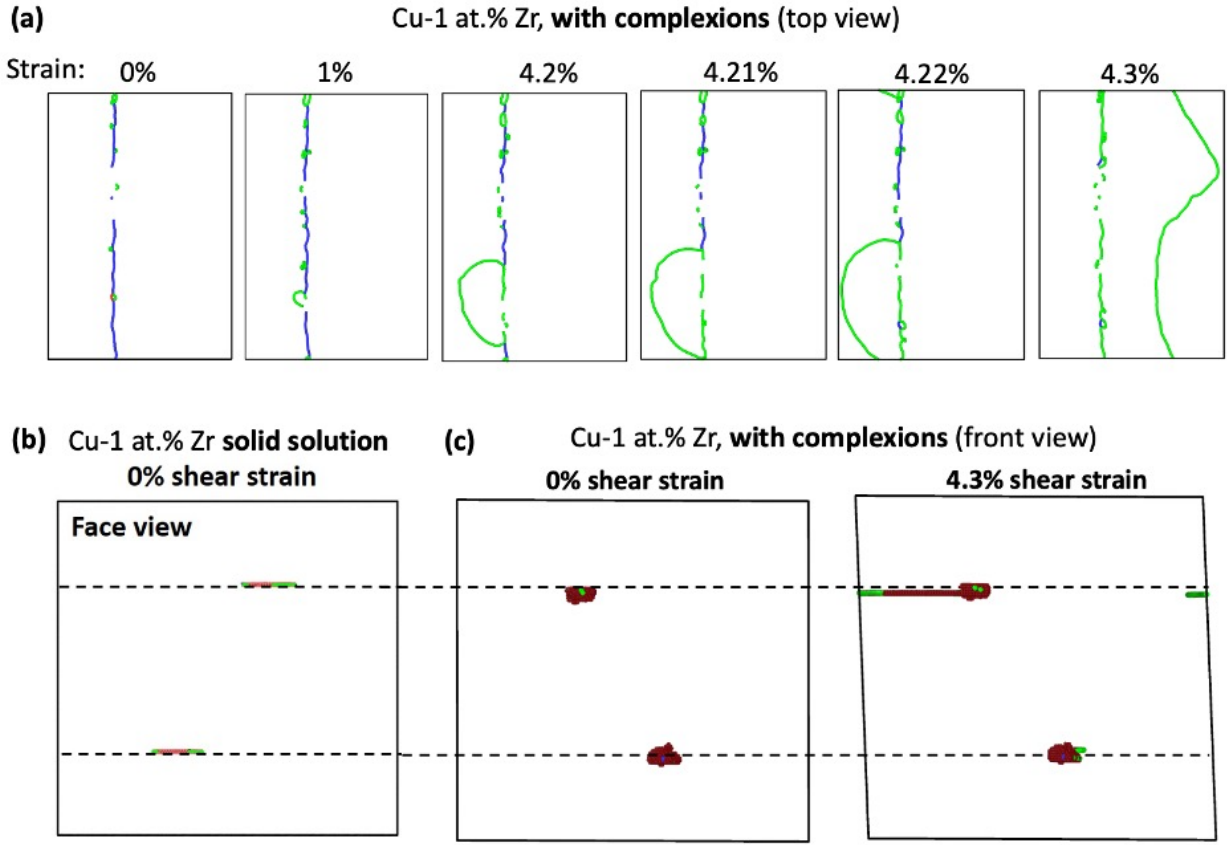

**Figure 9. (a) Top view of a stacking fault linear complexion in Cu-1 at.% Zr for increasing applied shear strain. A disconnected collection of defects is observed, with the nucleation of a leading partial occurring from a segment of this structure. Nucleation of this partial changes the overall dislocation character from full (blue) to partial (green). (b)-(c) A front view of the simulation cell, looking down the dislocation line direction. (b) Dislocation motion in the solid solution remains constrained to the initial slip planes (denoted by two dashed black lines), while (c) dislocation nucleation can occur in a different, parallel slip planes due to the delocalization of the dislocation core.**

## 4. Discussion

With an understanding of the fundamental dislocation mechanisms, we now return to the measured strength values reported in Figures 3(b) and (c). For all five alloy systems, the breakaway stresses of the linear complexion samples are significantly higher than those of the solid solution samples, demonstrating that complexions are much stronger obstacles to initial yield. Specifically, the linear complexions samples in the Ni-Al, Al-Zr, Al-Cu, and Al-Cu-Zr alloys have comparable normalized breakaway stresses in the range of 0.013-0.016 $G$, or 4-6 times higher than the comparable solid solutions. The comparison with the solid solutions can be understood by the fact that solutes are modest obstacles to slip that are distributed randomly in the solvent, while linear complexions are localized to the region near the dislocation. Having a stronger obstacle generated right at the defect results in more resistance to the initial dislocation motion. Although Cottrell atmospheres are also an example of localized segregation around the dislocations, local chemical ordering exists in the case of linear complexions and can modify the slip environment more strongly. For these four complexions, the original dislocation core eventually moves once it overcomes the local obstacles comprised of arrays of nanoparticles or platelets. While there is a similar strengthening effect, we note that the obstacle densities are not identical. The nanoparticle array linear complexions shown in Figure 1(b) for Ni-Al and Al-Zr more fully decorate the dislocation line than the platelet array complexions shown in Figure 1(c) for Al-Cu and Al-Cu-Zr. With less dense spacing yet a slightly stronger strengthening effect, one can conclude that the individual platelets provide stronger obstacles for the initial breakaway event than individual nanoparticles. This observation aligns with the mechanisms explored above, as the faceting of the dislocation line and the creation of stable edge character segments along the GP zone platelets provides additional well-defined obstacles to resist dislocation motion. Comparison of the Al-Cu complexions (only GP zone platelets) with the Al-Cu-Zr complexions (GP zone platelets plus $L1_2$ nanoparticles) shows that the addition of a secondary complexion type only gives a very minor strengthening increment. By far the highest breakaway stress is observed for the Cu-Zr stacking fault linear complexion specimens, which are 2.5-3 times stronger than the other linear complexion types. While the original dislocation can eventually move away from the other complexions, the requirement for dislocation nucleation results in a higher strength for this defect configuration. The breakaway stress corresponding to the nucleation of this Shockley partial is ~0.04G, or 57% of the ideal pure shear strength of Cu [56]. Also notable is the very low breakaway stress for the solid solution Cu-Zr, which means that the gap between complexion (highest value from entire dataset) and solid solution (lowest value from entire dataset) samples is particularly pronounced in this case, with a strengthening factor of 70× (Figure 3(b)).

The flow stresses can next be discussed for different sample types. Inspection of the atomistic configuration at the measured peak stress values from the solid solution samples shows that these points are primarily associated with the two dislocations in the simulation cell mutually interacting through their respective stress fields, and are therefore not a true measurement of the solid solution itself. In contrast, the peak stresses from the linear complexion samples correspond to atomic configurations where the dislocations re-enter the cell and interact again with the complexion structure. As such, we limit this discussion to the linear complexion samples. For the nanoparticle and platelet array linear complexions, the normalized peak stresses associated with plastic flow have similar magnitudes to the initial breakaway stress, due to the fact that the dislocation returns to its original location each time it travels through the periodic boundaries. The dislocations move around and away from the obstacles, but do not interact with them directly or alter the obstacle structure. Instead, the longer-range stress fields communicate. The dislocations become locked again upon re-entry and must overcome a similar energetic barrier to move. The normalized peak stresses for the Cu-Zr sample with stacking fault linear complexions are still the highest, but a significant decrease is observed from the breakaway stress. This observation can be explained by the fact that the specimen does not return to exactly the same configuration after dislocation nucleation. The complexion structure is altered by nucleation, as shown in Figure 9(a), meaning that subsequent nucleation events are easier. The stress-strain curves in Figure 2 shown this clearly, with noticeable softening after the initial yield event for the Cu-Zr linear complexion samples.

To fully explore the details of the dislocation restrictions associated with linear complexions, it is useful to compare against traditional strengthening mechanisms. Figure 10 provides schematic views and important details to describe the three main linear complexion types (right column), as well as their most comparable traditional strengthening mechanism (left column). Nanoparticle array linear complexions are compared to bulk precipitation strengthening (top row), platelet array linear complexions are compared to bulk GP zones (middle row), and stacking fault linear complexions are compared to heterogeneous nucleation of dislocations from grain boundaries (bottom row). Second phase precipitates (top left) behave as obstacles when they lie in a slip plane, acting to physically block the path of a dislocation. The most common deformation mechanism occurs when the dislocation cannot penetrate the particle and must bow around it with through a bowing mechanism, leaving behind a sessile Orowan loop that surrounds the precipitate and acts as debris (particle shearing can also occur for small, coherent precipitates). While this mechanism is well-known and is taught in textbooks (see, e.g., [71]), research continues on the topic to this day (see, e.g., Ref. [72]). Xu et al. [73] used concurrent atomistic-continuum simulations to study dislocation motion past an array of obstacles in Al, finding that cooperative dislocation bow-out on different length scales was an important mechanism. In contrast, the obstacles in a nanoparticle array linear complexion (top right) are either above or below the slip plane, but do not provide a physical barrier. Instead, the reduced stress field of the dislocation serves as a restraining force. Because of this altered orientation with respect to the slip plane, there is no debris left over after the breakaway event (or sheared particles, in the case of small coherent precipitates), leaving pristine collections of nanoparticles. Other dislocations moving through the material in later stages of plasticity could interact with these particles in a similar way as long as their stress fields overlap the particle location, providing a template of strong obstacles within the material.

Moving to the bulk platelet strengthening mechanism, GP zones (middle left) are found in many underaged precipitation strengthened alloys, being particularly common in Al-rich systems (see, e.g., [33-42]. These platelets remain coherent with the matrix and are extremely thin, often only a few atomic layers in thickness [42] and are therefore typically traversed directly by dislocations. This process leaves a sheared or "cut" GP zone in the microstructure. Singh and Warner [34] studied dislocation interactions with GP zones in Al-Cu using atomistic simulations, demonstrating that bowing can sometimes accompany this shearing mechanism. For example, for a 4.4 nm GP zone with 60° orientation with respect to the incoming dislocation, these authors reported that the leading partial dislocation cuts through the precipitate, damage the platelet, while the trailing partial bows around the feature, leaving debris. In all cases, damage or debris was left behind by the platelet-dislocation interaction. In contrast, for platelet array linear complexions, no damage occurs because the platelet is above or below the glide plane and is therefore bypassed. Turlo and Rupert [9] showed that these GP zones grow along {100} planes and have lower atomic volume than the matrix, so it is likely that the edge dislocation facets form to reduce the impact of their compressive stress field. Unpinning from these platelet array complexions is therefore associated with overcoming the local stress field interactions, rather than a physical obstacle, and no damage is left behind. The lack of damage and/or debris is common to both the nanoparticle and platelet array linear complexions, as is the targeted location of the obstacles. Both bulk second phase precipitates and bulk GP zones nucleate at random locations within the material, meaning that dislocations must travel some distance before interacting with the obstacles by chance. In contrast, complexion formation is driven by the stress field of the dislocations, meaning that the obstacles are born at the best possible location, ready to restrict motion immediately.

Heterogeneous dislocation nucleation marks the final traditional plasticity mechanism and is primarily active in materials with very few or no glissile dislocations, such as nanocrystalline metals [74], nanowires [75, 76], ultra-thin films [77], and well-annealing single crystals [78]. Using nanocrystalline metals as a well-studied point of comparison, the nanoscale grain interiors lack pre-existing dislocation content to sustain plastic flow, leading to the nucleation of dislocations from interfacial sites. Grain boundaries contains a network of dislocations with both lattice and non-lattice Burgers vector character, which Kuhr and Farkas [79] showed can either directly act as a source of defects to be pushed into the grain interior or serve as a local stress concentration for the nucleation of new lattice dislocations. These emitted dislocations are usually absorbed into the opposite grain boundary after traversing the grain [80], leaving no residual dislocation debris [81]. Stacking fault linear complexions also contain a disconnected network of dislocations as a result of the delocalized core, with these defects similarly acting as a dislocation nucleation/emission source at high applied stress. As dislocations cannot abruptly start or stop inside of a crystal, the defect is anchored on both sides of the nucleation site and an unzipping motion occurs as the dislocation moves away from the complexion (see Figure 9(a)). The complexion can also act as a dislocation sink or absorption site (see Figure 8), in this case once the dislocation has left and returned through the periodic boundary conditions. While this exact sequence is strongly influenced by our simulation geometry and boundary conditions, one can imagine a more realistic situation where a dislocation is nucleated from one stacking fault linear complexion and then encounters another further along the same slip plane. The growth of the stacking fault linear complexion out of the slip plane (Figure 9(c)) even means that a parallel but nearby slip plane can also host possible dislocation sinks.

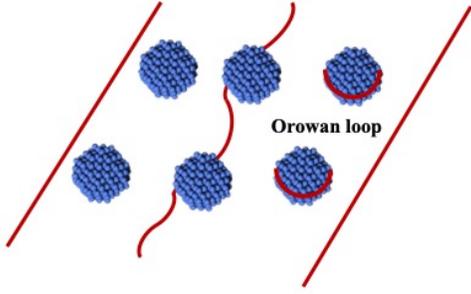

**Figure 10.** Comparison between dislocations mechanics in presence of traditional obstacles (left column) and linear complexions (right column). The top row shows particle-dislocation interactions, the middle row shows platelet-dislocation interactions, and the bottom row shows dislocation nucleation events.

While direct comparisons of the strengths measured here with experimental reports are difficult, due to the high strain rates used in MD and also due to the isolation of individual defects in this work, the relative strengthening effect associated with similar obstacles is worthy of discussion. The most important descriptor in future experiments is expected to be the fraction of glissile dislocations within the material that are decorated by linear complexions, as this metric will determine how many defects have heavily restricted motion. Even without a direct comparison, previous studies on the strengthening effect of bulk precipitates can provide a reference point for appreciating the impact of nanoparticle array and platelet array linear complexions. Many Ni-based alloys and superalloys have demonstrated outstanding mechanical properties in the presence of bulk $L1_2$ precipitates by classical precipitate strengthening [82-90]. For example, an improvement in the yield strength of up to 400% was reported in Ni-based superalloys owing to the presence of nanoscale, ordered $L1_2$-type precipitates, as compared to a similar alloy with only solid solution strengthening contributions [85]. A bimodal distribution of precipitate particles was observed in this alloy, with 70 nm particles present at a 30% volume fraction and 10 nm particles present at a 10% volume fraction. In another example, a hot-rolled and annealed Monel K500 containing γ' ($L1_2$) precipitates exhibited a 200% higher yield strength than Monel 400, which is the solid solution strengthened alloy of same composition [86]. In the case of Al-Cu alloys, precipitation strengthening was found to increase the yield strength up to a maximum of ~64% when compared to a solid solution sample of similar bulk composition [36, 41]. While the strengthening from bulk precipitates can be appreciable, the magnitude of the effect imparted by these features is strongly dependent on precipitate size and distribution and/or volume fraction. For example, Rodriguez et al. [36] performed a detailed study on the strengthening contribution of GP zones in naturally aged (293 K) Al-4 wt.% Cu samples. These materials had small GP zones of diameter $2.5 \pm 0.1$ nm and a volume fraction of $1.0 \pm 0.3$, which gave a CRSS of 88.1 MPa (a strength increase of 44%) compared to the solid solution sample. Overall, the examples given above show that bulk precipitates can strengthen to high levels, but only if high volume fractions of precipitates are added. The addition of only a few nanoscale obstacles generally results in modest increases in strength. Bulk precipitates are inefficient strengthening features because they rely on accidental dislocation-precipitate interactions as the dislocation moves through the lattice. Alternatively, nanoparticle and platelet arrays which are linear complexion counterparts of such bulk precipitates can give dramatic strengthening (400-600% higher than solid solutions in Figure 3(b)) by placing the obstacles immediately at the dislocations. Since linear complexions develop due to the dislocation's stress field, the obstacle to flow is at the exact right place and thus the probability of interaction is explicitly ensured.

Stacking fault linear complexions require a different class of materials for comparison, specifically those where dislocation nucleation is a key process. Two example types of materials where dislocation nucleation governs are nanocrystalline metals and nanowires, where grain boundaries and free surfaces often act as the nucleation sources. Both experiments and MD simulations have shown that dislocation emission becomes the dominant source of plasticity in nanocrystalline and nanowire samples, with the spacing of nucleation sites (i.e., grain size or nanowire diameter) being critical for determining the strength of these materials [91-97]. Roos et al. [94] and Bin Wu and coworkers [93] performed in situ transmission electron microscopy tensile tests on Au nanowires with diameters ranging from 30 to 300 nm. Richter et al. [96] similarly studied the strength of single crystal Cu nano whiskers for a diameter range of 75 to 300 nm. These studies revealed that yield strength is controlled by wire diameter. The Au nanowires show $3.5\pm1$ GPa yield strength for a 200 nm diameter wire and $5.6\pm1.4$ GPa yield strength for a 40 nm diameter wire [95]. The Cu whisker strength could approach the theoretical limit as wire diameter was reduced to ~100 nm [96]. Atomistic investigations of strengthening in Au nanowires reveal a similar inverse correlation between the strength and size of Au nanowires (see, e.g., [97]). Nanowires can therefore achieve great strengths by making it difficult to nucleate the first dislocation, but very fine wires with nanoscale diameters mean that the overall load bearing capacity of the material is small. In addition, synthesis of fine nanowires typically requires the use of novel techniques such as templated deposition (see, e.g., [98]) or vapor-liquid-solid growth techniques (see, e.g., [99]) which are extremely slow to produce appreciable material volumes. This manifests as limited wire dimensions along the growth axis, with lengths of 50 μm and 75 μm in the studies by Roos et al. [95] and Richter et al. [97], respectively. A similar deformation mechanism has also been observed in faulted nanowires, providing another comparison point for stacking fault linear complexions. Veerababu et al. [103] simulated the deformation of Cu nanowires decorated with stacking faults using MD, finding that plastic deformation was associated with the nucleation of stacking faults enclosed by leading partial dislocations from the nanowire surface.

The high strength of the nanocrystalline materials is also attributed to dislocation nucleation and emission, with grain boundaries acting as the source. Hugo et al. [100] and Kumar et al. [101] studied the deformation of nanocrystalline Ni films using in situ transmission electron microscopy, finding evidence of dislocation nucleation for grain sizes of ~20 nm. Nanocrystalline Pt films with an average grain size of 10 nm were also shown to experience dislocation nucleation from the grain boundaries during deformation [102]. The critical shear stresses required to nucleate partial dislocations were estimated to be ~2 GPa and ~12 GPa in grain sizes of 24 nm and 2 nm, respectively [102]. In another study, Chen et al. [91] reported the tensile strength of nanocrystalline Cu can become as high as ~1 GPa for an average grain size of 10 nm [91]. Like nanowires, the synthesis of nanocrystalline metals typically requires specialized processing routes, such high-energy ball mill, melt spinning, or vapor deposition, which severely limit the specimen size that can be fabricated. For instance, the Pt nanocrystalline thin film studied by Li et al. were synthesized by magnetron sputtering onto a (001)-oriented NaCl single crystal substrate, resulting in a final film thickness of only ~10 nm [102]. When compared to nanocrystalline metals and nanowires, stacking fault linear complexions also require dislocation nucleation yet can provide comparable strengthening levels without the above mentioned size limitations or the requirement of sophisticated fabrication routes.

## 5. Conclusions

In this study, atomistic simulations were used to probe the effect of three different classes of linear complexions recently predicted for face-centered cubic alloys on the plasticity, dislocation mechanics, and strengthening of these materials. Important findings from the current work include:

- The breakaway and peak stresses in all of the samples with linear complexions are significantly higher than the corresponding solid solution configurations, with the largest difference being 70× for Cu-Zr. By seeding strong obstacles to dislocation motion directly near the defects that control plasticity, linear complexions have the potential to enable engineering alloys with extreme strength.

- A comparison of the normalized breakaway stress shows that each type of linear complexion has a unique effect on the dislocation core in terms of both the pinning phenomenon and also the level of predicted strengthening.

- Nanoparticle array linear complexions pin the dislocations they are born at, but do not alter the dislocation core structure or obstruct the original glide plane in any away. The partials unpin from the nanoparticle arrays either as a pair or one-by-one, causing the dislocations to bow out. Detailed study of the dislocation motion and stress-strain trends indicate that the pinning power depends on the size and local arrangement of precipitate arrays along the dislocation line.
- Platelet array linear complexions present themselves as out-of-plane GP zones that nucleate along the partial dislocation lines. Again, the pinning power of the obstacles depends on the spatial distribution and size of the individual platelets, but appears to be stronger than the nanoparticle array configuration for a given complexion density.
- Stacking fault linear complexions delocalize the original dislocation core into a network of disconnected dislocations, reminiscent of grain boundary dislocations. The nucleation of a partial dislocation from the reduced core marks the onset of plasticity in this sample. As the nucleation event requires a much higher stress, approaching the theoretical shear stress of pure Cu, compared to the prior unpinning mechanisms observed for other complexion types, the stacking fault linear complexions offer the greatest promise for extreme strength.

Linear complexions are a new category of engineered defect structures, demonstrating the potential to directly manipulate dislocations and control plasticity in face-centered cubic alloys. These strengthening features are different from traditional strengthening mechanisms in a number of important ways, most notable being the lack of obstacle damage and the templating of the reinforcing features immediately at the dislocations. As a whole, the results presented in this work suggest that controlled incorporated of collections of nano-sized complexions may allow for new strengthening pathways. Thermomechanical processing routes that target the formation of linear complexions, and ultimately the development of plasticity models that incorporate complexion mechanics, would be fruitful targets for future study.

**Declaration of Competing interests**

The authors declare that they have no known competing financial interests or personal relationships that could have appeared to influence the work reported in this paper.

**Acknowledgments**

Research was sponsored by the Army Research Office under Grant Number W911NF-21-1-0288. The views and conclusions contained in this document are those of the authors and should not be interpreted as representing the official policies, either expressed or implied, of the Army Research Office or the U.S. Government. The U.S. Government is authorized to reproduce and distribute reprints for Government purposes notwithstanding any copyright notation herein.

**Data Availability**

The data that support the findings of this study are available within the article.